\documentclass[12pt]{article}
\pdfoutput=1
\usepackage{graphics, color}
\usepackage{graphicx}
\usepackage{amsmath}
\usepackage{amssymb}
\usepackage{amsfonts}
\usepackage{tensor}
\usepackage[usenames,dvipsnames]{xcolor}
\usepackage[normalem]{ulem}
\usepackage[bottom]{footmisc}

\definecolor{orange}{rgb}{1,0.5,0}
\definecolor{grey}{rgb}{.5,.5,.5}
\definecolor{bluegreen}{rgb}{0,.5,.5}
\definecolor{darkgreen}{rgb}{0,.5,0}

\def\gsim{\, \rlap{$>$}{\lower 1.1ex\hbox{$\sim$}}\,}
\def\lsim{\, \rlap{$<$}{\lower 1.1ex\hbox{$\sim$}}\,}
\newcommand{\be}{\begin{equation}}
\newcommand{\ee}{\end{equation}}
\newcommand{\bea}{\begin{eqnarray}}
\newcommand{\eea}{\end{eqnarray}}

\let\oldbibliography\thebibliography
\renewcommand{\thebibliography}[1]{%
  \oldbibliography{#1}%
  \setlength{\itemsep}{0pt}%
}
\begin{document}

%Title page

\begin{titlepage}
\bigskip
\bigskip\bigskip\bigskip
%\centerline{\Large \bf Something something something}
\centerline{\Large \bf Why trust  a theory?  Some further remarks (part 1).}
\bigskip

%\centerline{\large \bf An apologia for string theory and the multiverse\footnote{It means the opposite of what it sounds like.}}

\bigskip\bigskip\bigskip
\bigskip\bigskip\bigskip

 \centerline
 {\bf Joseph Polchinski\footnote{\tt joep@kitp.ucsb.edu }}
 \medskip
 
 \centerline{\em Kavli Institute for Theoretical Physics}
\centerline{\em University of California}
\centerline{\em Santa Barbara, CA 93106-4030 USA}

\bigskip\bigskip\bigskip
%ABSTRACT

\begin{abstract}
I expand on some ideas from my recent review ``String theory to the rescue.''  I discuss my use of Bayesian reasoning.  I argue that it can be useful but that it is very far from the central point of the discussion.  I then review my own personal history with the multiverse.  Finally I respond to some of the criticisms of Ellis and Silk, which initiated this interesting discussion.

Prepared for the meeting ``Why Trust a Theory? Reconsidering Scientific Methodology in Light of Modern Physics,''  Munich, Dec. 7-9, 2015.
\end{abstract}
\end{titlepage}

\baselineskip = 16pt
\tableofcontents

\baselineskip = 18pt

\setcounter{footnote}{0}

%\tableofcontents

\section{Introduction}

The meeting ``Why Trust a Theory? Reconsidering Scientific Methodology in Light of Modern Physics,''  which took place at the Ludwig Maximilian University  Munich, Dec. 7-9 2015, was for me a great opportunity to think in a broad way about where we stand in the search for a theory of fundamental physics.   My thoughts are now posted at~\cite{Polchinski:2015pzt}.  

In this followup discussion I have two goals.  The first is to expand on some of the ideas for the first talk, and also to emphasize some aspects of the discussion that I believe need more attention.  As the only scientific representative of the multiverse at that meeting, a major goal for me was to explain why I believe  with a fairly high degree of probability that this is the nature of our universe.  But an equally important goal was to explain why, contrary to what many believe, string theory has been a tremendous success, and remains a great and inspiring hope for our ultimate success in understanding fundamental physics.  For natural reasons, certain aspects of the multiverse discussion have gotten the greatest \ initial reaction.   I  will explain why some things that have been most widely discussed are actually the least important, while other things are much more central.

The second goal is to respond to some of the other speakers.  I was unexpectedly unable to attend the actual meeting.  The videos from the other talks will soon be posted, and I look forward to seeing them.  However, because they are not yet available, this followup will be presented in a somewhat unusual way.   The initial post will contain only part one.  Part two will be added when this becomes possible.

To summarize the main theme of~\cite{Polchinski:2015pzt}, nature has presented us with two great challenges.  The first is that the Planck length, the fundamental length scale at which quantum mechanics and relativity come together, is incredibly small.  The second is the possibility that the constants of nature that we are working so hard to measure at the LHC and elsewhere are environmental.  If so then they are very different from the constants of nature that we have usually dealt with, such as those in the theories of the weak or strong interaction.  There, from a very small number of inputs from the Weinberg-Salam theory or QCD, one can predict with very high precision hundreds of measurable properties of elementary particles.  This is `normal physics.'  But if some of the constants of nature, like the masses of the quarks and leptons, are environmental, it means that these are much more loosely connected to the fundamental mathematics that underlies physics.  There is an essential randomness to them, not unlike the weather.  This is not `normal physics,' and it is not what anyone wants, but unfortunately there is significant evidence that this is the case.

These difficulties seem often to be presented as failures of string theory, as though trying a different idea will somehow make the Planck length larger, or force that environmental property not to hold.  But these are challenges for humankind as a whole, at least for those of us who care about the fundamental laws of nature.  And the conclusion that I came to is that they are not insuperable, that we will succeed in figuring things out, and that string theory is a large part of the answer.

Both \S2 and \S3 are largely concerned with the multiverse.  In \S2 I explain some statistical issues that have gotten wide attention.  In \S3 I review my own long scientific interaction with the idea of a multiverse, because it provides a good framework for explaining how and why we have come to this idea. In \S4 I  briefly discuss some of the criticism of Ellis and Silk, which initiated this interesting discussion.  A more complete discussion of related issues appears on my web site at http://www.kitp.ucsb.edu/why-trust-a-theory-reconsidering-scientific-methodology-in-light-of-modern-physics.

\section{It's not about the Bayes.  It's about the physics.}

In \S3 of~\cite{Polchinski:2015pzt} I made an important distinction between the 5-sigma or 99.9999\% probability for the Higgs boson to exist, and a 2-sigma or 94\% probability for the multiverse to exist.  For a particle physicist, the difference between such numbers is a familiar thing.  However, I have seen that for members of the public, for many cynics, and even for some of the philosophers for whom the talk was written, this distinction is not so obvious, and so I will explain.  The 5-sigma standard for the Higgs has been widely discussed in many blogs and other public sources, so I will not expand on it here.  It is the meaning of the 2 sigma that I want to explain.

I am coming at this as an active scientist trying to solve a scientific problem, in my case the theory of quantum gravity.  Every day I need to make my best estimate of where to put my effort.  This of course is not unique to the problem that I am working on.  For example, every experimental or theoretical particle physicist, confronted with evidence for some new particle or dark matter observation, wants to know how strong is the evidence, what is the sigma, so they know how much effort to apply to the problem.

There is inevitably  a significant amount of personal taste and judgement, and there is no right answer at this level for a `correct' value of sigma.   Science works better if different people try different assumptions.  For my own problem,  it is essential to factor in what ideas I and others have already tried.  This has been criticized in some quarters, and indeed it might not be proper for a discovery measurement like the Higgs.  But for my purpose it is obviously necessary.

In the end, my estimate simply came down to combining four factors of two.  This led to the probability 94\% that the multiverse exists.  If I had quoted this in binary, probability 0.1111, the  scientific content would have been exactly the same but it would have led to much less merriment.  

My estimate uses a kind of analysis known as Bayesian statistics.  There has recently been much discussion of this method, partly in response to the Munich meeting.  So let me say very clearly: Bayesian analysis is not the point.  It is not even one percent of the point.  Every word spent on this subject is a wasted word, in fact it has negative value because it distracts attention from the real point, which is of course the physics.  And I can blame myself for this, because I chose to frame the problem in this way.  So let  me summarize what the relevant physics is.  In the next section I give a longer discussion of some of the ideas, and my own personal history with them.

The main issue is the vacuum energy, the cosmological constant.  This goes back a long way, to Pauli, who argued in 1920 that the zero point energy of the radiation field would cause the radius of the universe to be no larger than the distance to the moon~\cite{Straumann:2002tv}.  In the ensuing 96 years, there has been essentially no progress on this problem using `normal' science. The difficulty has in fact been sharpened: the vacuum energy is too large by a factor of $10^{120}$, though this can be reduced to $10^{60}$ in theories with low energy supersymmetry.

So normal science has a problem: it gives a universe that does not  look like ours.  But if the cosmological constant is environmental --- if it depends on space or time --- then suddenly we get a universe like ours, where observers find themselves in a large universe with small curvature.

The environmental theory comes with a great surprise.  In the context of normal science, it was almost universally assumed that the actual value of the cosmological constant was exactly zero, and this was what we were trying to explain.  But in the environmental theory there is no reason for it to be zero, and in fact the value zero would be infinitely unlikely.  The prediction is just that the cosmological constant be small enough (more on this  later).  And of course this is what was discovered in 1998~\cite{Riess:1998cb,Perlmutter:1998np}, to the great surprise of many --- but not those who had been following Weinberg's environmental approach.

Finally, string theory, which I credit as an enormous success for reasons reviewed in \S2 of~\cite{Polchinski:2015pzt}, leads to environmental physics, the string landscape.

In all, there are four arguments here for the multiverse: the failure of conventional methods for understanding why the cosmological constant is not large, the success of the environmental theories for doing so, the successful prediction of the nonzero cosmological constant, and the string landscape.  This is the physics, with no reference to Bayes.

Returning to Bayes, I chose to package these four numbers each as a factor of two, a nice mnemonic for the four kinds of evidence, leading to the 94\%.  Perhaps it was silly to do so, but expressing things in terms of numbers is good, provided this is taken in the correct spirit.  That is, we are not talking about discovery mode like the Higgs.  We are trying to make a best estimate of our current state of knowledge as guidance to how to move forward.  Of course those who choose to belittle will always do so, but these are not the ones to listen to.

Obviously there is no precision to our number.  Andrei Linde would argue for a larger number~\cite{Linde:2015edk}, taking into additional cosmological evidence beyond the cosmological constant.  I would have come up with a larger number if I had given more weight to my long history trying to understand the cosmological constant by normal methods, both my own ideas and those of others.  But I thought it important to factor in possible unknown unknowns.  While I believe that the road to quantum gravity will go through string theory, there may still be new and game-changing concepts along the way. 

\section{The multiverse and me}

I have sometimes seen the assertion that the multiverse is largely the resort of older theorists who, having run out of ideas or time, are willing to give up the `usual' process of science and use the multiverse as an excuse~\cite{4g}.  In my own experience much the opposite is true.  My own interaction with the multiverse began at the age of 34, and I think that this is much more typical.  In this section I will review my own personal associations with the multiverse, because I believe that it is instructive in a number of ways.  I will also discuss some of the experiences of two scientists with whom I have been closely associated, Steven Weinberg and Raphael Bousso.  Other important perspectives would include Andrei Linde~\cite{Linde:2015edk}, Tom Banks~\cite{Banks:1984cw} (though he is no longer an advocate of these ideas), Bert Schellekens~\cite{Schellekens:2013bpa}, and Lenny Susskind~\cite{Susskind:2003kw}.

I began to think about the problem of the cosmological constant as a postdoc, around 1980.  The then-new idea of inflation had emphasized the importance of the vacuum energy, and of the connections between particle physics and cosmology more generally.  The presumed goal was to understand why the value of the cosmological constant was exactly zero.  I went through the usual attempts and tried to find some new wrinkles, a process that began with Pauli and has certainly been repeated by countless others  over the years.  I failed to explain why the cosmological constant would vanish.\footnote{As an aside, there is a certain collection of ideas that have some aspects of the environmental approach, in that the value of the cosmological constant can change in time or in different branches of the wavefunction~\cite{Banks:1984cw,Hawking:1984hk,Abbott:1984qf,Brown:1988kg,Coleman:1988tj}.  However, for these the cosmological constant changes in a rather systematic way, that is amenable to `normal science.'  By contrast, in the environmental approach the calculation of the cosmological constant  is a problem of a particularly hard sort, NP hard~\cite{Denef:2006ad}.}  I added this to my list of the most important questions to keep in mind, believing that we would one day solve it and that it would be a crucial clue in understanding fundamental physics.

I eventually got a good opportunity to review my attempts at the problem~\cite{Polchinski:2006gy}.  Bousso has presented a similar review~\cite{Bousso:2007gp}.  Gerard 't Hooft had proposed for many years  to write a review ``100 Solutions to the Cosmological Constant Problem, and Why They are Wrong."  He never did so, but decided that it would be a good exercise for his grad student~\cite{Nobbenhuis:2004wn}.

In 1988, Weinberg similarly set out to review the possible solutions to the cosmological constant problem, but he approached it in a particularly broad and original way~\cite{Weinberg:1988cp}.  Of course Weinberg is a great scientist, but there are two qualities he has that I find distinctively striking, and that are special to him.  As science progresses, it often happens that certain assumptions will become widely accepted without really being properly inspected.  Eventually someone asks the question, ``What if B is true instead of A,'' and they may find that B is a perfectly good possibility, and a new set of ideas emerges.  Weinberg of course is not the only person who asks such questions, though he often is the one, but when he does so he does it in a way that is distinctive and powerful.  Most of us will say, ``Let us see what happens if we make a model in which B is true instead of A.  But this looks hard, so we'll also assume C and D so we can get an answer.''  Weinberg does not do this.  As much as possible he tries to ask, ``If B is true instead of A, then what is the most general set of consequences that one can draw?''  Generally one cannot make progress without some additional assumptions, but his effort is always to make these minimal, and most importantly to state as precisely as possible any such extra assumptions being made.  It is an approach that takes great clarity of mind, to formulate such a general question in a way that can be usefully answered, and great integrity not to take shortcuts.  It is an approach that Weinberg has applied many times, usually in more ordinary areas of science, to great effect.

The second and related quality is that he takes an unusually broad  and independent point of view as to for what the possibilities for B might be.  One example is asymptotic safety, the possibility that the short distance problem of quantum gravity is solved, not by new physics like string theory, but by a nontrivial UV fixed point of the renormalization group~\cite{Wsafe}.  A second example, the one to which he turned in his 1988 review, is the possibility that the cosmological constant might be environmental, varying from place to place.   Other theorists had already begun to explore the idea~\cite{DavUnr.1981,Linde:1984ir,Sakharov:1984ir,Banks:1984cw}, but Weinberg brought to it his unique  strength.  He argued that most places will be uninteresting, with space either expanding so rapidly as to be essentially empty, or contracting to a singularity long before anything interesting can happen.  But there will be a sweet spot in between, where spacetime looks much like what we see around us.  Here galaxies would form, and life would presumably follow by the usual processes, and eventually creatures would form who could do cosmological measurements.  And because this can only happen in the sweet spot, they will find a value for the cosmological constant that is extremely small, just as we find.

I saw Weinberg's work at first hand, because at the time I was a colleague of his at the University of Texas at Austin.  The effect of his analysis on me was profoundly upsetting.  If correct, he had solved the problem that I had failed to solve, but he had done so in a way that took away some of our most important tools.  I had hoped to make progress in fundamental physics by finding the symmetry or dynamics that would explain why the cosmological constant is exactly zero.  But now it was no longer zero.  Even worse, it is not a number that we can calculate at all, because it takes different values in different places, a vast number of possible values running over different solutions.

So I very much wanted Weinberg to be wrong, and that the cosmological constant would turn out to be exactly zero.  Unfortunately there was already some evidence that it was not.  There was the age problem, the fact there seemed to be stars that were older than the universe.  And there was the missing mass, the fact that the total amount of matter that could be found seemed to account for only 30\% of the energy budget of the universe.  Neither of these requires  the cosmological constant, but the simplest way to make it all balance would be if the cosmological constant were nonzero.  There was also a sign that Weinberg might be wrong.  His estimate for the cosmological constant came out a factor of $10^2$ or $10^3$ too large.  But missing by $10^3$ is a lot better than missing by $10^{60}$, and one could see that by changing some of the details of Weinberg's argument one might end up with a much closer number.  So I spent the next ten years hoping that the age problem and the missing mass problem would go away, but they did not.  And in 1998 the cosmological constant was confirmed as the correct answer.

Fortunately, in 1998 there was something much more interesting for string theorists to be thinking about, so the cosmological constant, important as it was, sat on the back burner for a while.  The second superstring revolution began in 1995.  Over a period of four years, we discovered dualities of quantum field theories, dualities of string theories, duality between quantum field theories and string theories (that is, AdS/CFT), D-branes, Matrix theory, and a quantitative understanding of  black hole entropy.  Our understanding of the basic structure of fundamental theory had moved forward in a way that seems almost miraculous.  One cannot compare this to something as profound as the discovery of quantum mechanics, or of calculus, but I think that there are  few points in the history of science when our depth of understanding moved forward so rapidly, and I feel fortunate to have been part of it.

I think that the capabilities that we had before and after the second superstring revolution are much like what the rest of the world experienced before and after the world wide web.  If I had to choose only one of these, I know which it would be.

Much of the initial analysis during the second superstring revolution focused on states with some amount of supersymmetry, because these new tools that we were learning are most tractable in this setting.  But all of these ideas apply to the full quantum theory, states of all kinds.  By 1999, the principles had been understood, and it was time to think about the implications for more physical systems.

Fortunately, Raphael Bousso was then visiting from Stanford, and both he and I had a strong sense that string theory as now understood provided the microscopic theory that was needed in order for Weinberg's environmental theory to work.  In his spirit of minimal assumptions, he had been able to formulate the problem in a way that did not require knowing the actual microscopic theory.  We had the perfect combination of skills, my understanding of the dynamics of strings and Raphael's understanding of cosmology.  We understood why many attempts to find a microscopic model of Weinberg's theory had failed, and we could see how it all worked in string theory.  We were quickly able to make a simple but convincing model that showed that string theory was very likely lead to exactly the kind of environmental theory that Weinberg had anticipated~\cite{Bousso:2000xa}.  I say quickly, but there was a significant delay due to  my own personal misgivings with the landscape, something that I will say more below.  But we published our model, and other string theorists began to show in detail that this is actually how string theory does work~\cite{Silverstein:2001xn,Kachru:2003aw}.

My misgivings with Weinberg's solution had not ended, in fact they had gotten worse.  Planck had taught us that because his scale was so small, the possible measurements of fundamental physics would be very limited.  Now Weinberg, with Raphael and my connivance, were saying that even these few things were actually largely random numbers.  We had gotten far, but it seemed that we would never get farther.  The Bayesian estimate was saying that there was a 94\% probability that our clues are just random numbers.  What then could I work on?
Shortly before the 1998 discovery of the cosmological constant, I told Sean Carroll that if the cosmological constant were found he could have my office, because physics would be over.

In fact, even when Raphael and I wrote our paper, I had pushed him to remove some of the most anthropic parts of the discussion.  It was not that I thought that they were wrong, we were in perfect agreement about the physics, but I felt that it was too discouraging.  In effect we would be telling  the experimentalists, who were spending billions of dollars and countless human-years of brilliant scientists, were essentially measuring random numbers, and I did not want to be the bearer of bad news.  But for the most part Raphael's point of view prevailed: we both agreed clearly on what the science was saying, and in the end one must be true to the science.  I am not by nature a radical, but seem to have become one  both with the multiverse and with the black hole firewall~\cite{Almheiri:2012rt} just by following ideas to their logical conclusions.  In both cases it helped to have brilliant and bold young collaborators, Raphael in the one case and Don Marolf in the other, who drove me to suppress my natural conservatism.

Still my anxiety grew, until eventually I needed serious help.  So you can say quite literally that the multiverse drove me to the psychiatrist.\footnote{In truth, I should have gotten help for anxiety sooner, and for more general reasons.  One should not be reluctant to seek help.}

In the end I have come to a much more positive point of view, as expressed in~\cite{Polchinski:2015pzt}.  We are looking after all for the theory of everything.  Actually, this is an expression that I dislike, and never use, because `everything' is both arrogant and ill-defined.  But I will use it on this one occasion to refer to the set of ideas that we are thinking about in quantum gravity, plus possibly some more that will turn out to be part of the story.  There will not be a unique route to the theory of everything.  If one route becomes more difficult, another will work.  It will likely require clues from many directions - particle physics, cosmology, certainly string theory, and likely other areas as well, but we will succeed.

\section{Some critics}

\subsection{George Ellis and Joseph Silk}

In late 2014, George Ellis and Joseph Silk published in Nature a criticism of string theory and of the multiverse --- indeed, an argument that it is not even science.  They argued that the issue was so important that a conference should be convened to discuss it.  I am very grateful that they followed through on this, because it is  indeed a vital issue, and it has given me a venue to present a much more positive and inspiring point of view.  I wish that I had done so long ago.

I would like to make a few comments on some specifics of Ellis and Silk.
One of the arguments that underlies their discussion is that physics is simply getting too hard, so we will never get the answer.  Indeed this is a logical possibility, and I worry about it.  But the lesson of Planck is that we are working on a problem that is very hard, and so we should think about this on a very long time scale.  We have been working on Planck's problem for 116 years and have made enormous progress.  It is far too early to say that we have reached the end of our rope.  It is a race between finding the key clues and running out of data and ideas, and I very much expect that we (humankind) will win.

A second disagreement is the claim of Ellis and Silk that unimodular gravity is a satisfactory alternative theory of the cosmological constant~\cite{Ellis:2013eqs}.  Unimodular gravity is Einstein's theory with one fewer equation, the one with the cosmological constant.  But the value of the cosmological constant matters, it affects physics, so we need some way to get that last piece.  In unimodular gravity, effectively the value of the cosmological constant is assumed to be measured, and whatever value we get is taken as the missing piece.  One gives up on trying to ever understand the particular value, which one finds to be around $7 \times 10^{-27}$ kg m$^3$.  Why is the value that one finds not the zero that Einstein expected?  Why is it not the $10^{60}$ larger that Pauli expected?  These simply cannot be explained.  This does not seem to me a satisfactory way for theory to work.  Moreover, it seems to me that it is {\it less} predictive than the multiverse.  Why the cosmological constant  is not enormous cannot be explained in unimodular gravity, nor can it be explained why it has a small nonzero value.  In contrast, the multiverse explains both.  So I would think that Ellis and Silk would regard the multiverse as a better scientific theory, but they have a different take.

However, like Ellis and Silk, I am trouble by the expression ``non-empirical theory confirmation.''  For me, it may just be that I have not understood clearly what is meant by this.   If someone is interpreting the rules in such a way that string theory is already at 5 sigma level, discovery, I cannot agree.   How could I, when we do not even know the full form of the theory?  And what is the rush?  Again, long ago Planck taught us that it will take a long time.  

But I think Dawid means something different, and more consistent with how I see things~\cite{Dawid:2013maa}.  In particular, he seems to be using the same kind of evidence that I argue for~\cite{Polchinski:2015pzt}.  Coming back to the theme of sigma values, I counted five kinds of evidence in \S2 and one more in \S3.  If I just make the same crude counting of a factor of 2 for each, I end up with probability 98.5 for string theory to be correct.  But here, where most of the focus is on quantum theory rather than the multiverse, I am on grounds that I understand better, and I would assign a higher level of confidence, something over 3 sigma.    

But again, these statistical games are fun, but they are not the real point.  There is some fascinating physics to explore right now, this is what we want to be doing, and it may transmute our understanding in ways that we cannot anticipate.

\section*{Acknowledgments}

I think Raphael Bousso, Melody McLaren, and Bill Zajc for advice and extensive proofreading.  
The work of J.P. was supported by NSF Grant PHY11-25915 (academic year) and PHY13-16748 (summer).

\end{document}